\documentclass{article}
\usepackage{graphicx}
\usepackage{amsmath}

\input{tcilatex}

\begin{document}

\title{Different Particle Sizes Obtained from Static and Dynamic Laser Light
Scattering }
\author{Yong Sun}
\maketitle

\begin{abstract}
Detailed investigation of static and dynamic laser light scattering has been
attempted in this work both theoretically and experimentally based on dilute
water dispersions of two different homogenous spherical particles,
polystyrene latexes and poly($N$-isopropylacrylamide) microgels. When
Rayleigh-Gans-Debye approximation is valid, a new radius $R_{s}$, referred
to as a static radius, can be obtained from the static light scattering $%
\left( SLS\right) $. If the absolute magnitude of the scattered intensity
and some constants that are related to the instrument and samples are known,
the average molar mass for large particles can be measured. The size
information obtained from SLS is purely related to the optical properties of
particles, i.e., to $R_{s}$ and its distribution $G\left( R_{s}\right) $.
The size information obtained from dynamic light scattering $\left(
DLS\right) $ is more complicated, the size distribution of which is a
composite distribution that is not only related to the optical properties of
particles, but also related to the hydrodynamic properties and the
scattering vector. Strictly speaking, an apparent hydrodynamic radius \ $%
R_{h,app}$ is a composite size obtained from averaging the term $\exp \left(
-q^{2}D\tau \right) $ in the static size distribution $G\left( R_{s}\right) $%
, with the weight $R_{s}^{6}P\left( q,R_{s}\right) $ that is also a function
of both $R_{s}$ and the scattering vector $q$.
\end{abstract}

For colloidal dispersion\ systems, the light scattering is a widely used
technique to measure the sizes of particles. During the last few decades,
the standard method of cumulant is used to obtain the apparent hydrodynamic
radius $R_{h,app}$ and its distribution $G\left( R_{h,app}\right) $ of
particles from the normalized time auto-correlation function of the
scattered light intensity $g^{\left( 2\right) }\left( \tau \right) $ $\left[
1-5\right] $ with the assistance of the Einstein-Stokes relation.\ The
treatment of the static light scattering $\left( SLS\right) $ spectroscopy
is simplified to\ the Zimm plot, Berry plot or Guinier plot etc. to get the
root mean-square radius of gyration $\left\langle R_{g}^{2}\right\rangle
^{1/2}$ and the molar mass of particles provided that the particle sizes are
small $\left[ 4,6,7\right] $.

The standard DLS techniques are not suited to accurately detect the
poly-dispersities of less than about 10\%. Pusey and van Megen $\left[ 7%
\right] $ proposed a method to detect small poly-dispersities when the
Rayleigh-Gans-Debye $\left( RGD\right) $ approximation is valid, measuring
the dependence of the effective diffusion coefficient obtained from the
initial slope of the correlation function with respect to the scattering
angle. By definition, the effective diffusion coefficient is an
intensity-weighted average diffusion coefficient. Meanwhile, when the
standard method of cumulant was used, from the fit results of our samples,
the values of second moment $\mu _{2}$ can have large differences even if
the experimental data were measured under the totally same conditions. More
importantly, the fit values of $\mu _{2}$ are often negative. It is a big
contradiction with its definition and lets the second moment $\mu _{2}$ lose
its physical meaning.

How the size distributions of particles were obtained directly from the SLS
data has been researched by a few authors $\left[ 9,10\right] $. In general,
people think that the radius obtained using the SLS technique is the same as
that obtained using the DLS technique if the effects of the different
weights of particle sizes and the scattering vector are considered. From our
results, a new radius can be obtained from the SLS data, namely static
radius $R_{s}$. For simplicity, in this paper, the homogenous spherical
particles are considered and the RGD approximation is thought to be valid.
The average scattered light intensity of a dilute non-interacting
poly-disperse system in unit volume can be obtained for vertically polarized
light

\begin{equation}
\frac{I_{s}}{I_{inc}}=\frac{4\pi ^{2}\sin ^{2}\theta _{1}n_{s}^{2}\left( 
\frac{dn}{dc}\right) _{c=0}^{2}c}{\lambda ^{4}r^{2}}\frac{4\pi \rho }{3}%
\frac{\int_{0}^{\infty }R_{s}^{6}P\left( q,R_{s}\right) G\left( R_{s}\right)
dR_{s}}{\int_{0}^{\infty }R_{s}^{3}G\left( R_{s}\right) dR_{s}},
\label{main}
\end{equation}
where $\theta _{1}$ is the angle between the polarization of the incident
electric field and the propagation direction of the scattered field, $c$ is
the mass concentration of particles, $r$ is the distance between the
scattering particle and the point of the intensity measurement, $\rho $ is
the density of the particles, $I_{inc}$ is the incident light intensity, $%
I_{s}$ is the intensity of the scattered light that reaches the detector, $%
R_{s}$ is\ the static radius of a particle,$\ q=\frac{4\pi }{\lambda }%
n_{s}\sin \frac{\theta }{2}$ is the scattering vector, $\lambda $\ is the
wavelength of the incident light in vacuo, $n_{s}$\ is the solvent
refractive index,$\ \theta $\ is the scattering angle, $G\left( R_{s}\right) 
$ is the number distribution and $P\left( q,R_{s}\right) $ is the form
factor of homogeneous spherical particles

\begin{equation}
P\left( q,R_{s}\right) =\frac{9}{q^{6}R_{s}^{6}}\left( \sin \left(
qR_{s}\right) -qR_{s}\cos \left( qR_{s}\right) \right) ^{2}.
\end{equation}
The average molar mass is defined as

\begin{equation}
\left\langle M\right\rangle =\frac{4\pi \rho }{3}N_{0}\int_{0}^{\infty
}R_{s}^{3}G\left( R_{s}\right) dR_{s},
\end{equation}
where $N_{0}$ represents Avogadro's number.

Comparing with the Zimm plot analysis, the mean square radius of gyration
for a poly-disperse system is

\begin{equation}
\left\langle R_{g}^{2}\right\rangle _{Zimm}=\frac{3\int_{0}^{\infty
}R_{s}^{8}G\left( R_{s}\right) dR_{s}}{5\int_{0}^{\infty }R_{s}^{6}G\left(
R_{s}\right) dR_{s}},  \label{RG}
\end{equation}
here $G\left( R_{s}\right) $ is chosen to be a Gaussian distribution

\begin{equation}
G\left( R_{s};\left\langle R_{s}\right\rangle ,\sigma \right) =\frac{1}{%
\sigma \sqrt{2\pi }}\exp \left( -\frac{1}{2}\left( \frac{R_{s}-\left\langle
R_{s}\right\rangle }{\sigma }\right) ^{2}\right) ,
\end{equation}
where $\left\langle R_{s}\right\rangle $ is the mean static radius and $%
\sigma $ is the standard deviation relative to the mean static radius.

Two standard polystyrene latex samples from Interfacial Dynamics Corporation
(Portland, Oregon) were used in SLS and DLS measurements. One is the sulfate
white polystyrene latex with a normalized mean diameter of 67 nm and the
other is the surfactant-free sulfate white polystyrene latex of 110 nm, as
shown in Table 1. All values were provided by the supplier as obtained using
Transmission Electron Microscopy $\left( TEM\right) $ technique. The $Latex$
1 was diluted for light scattering to weight factor of $1.02\times 10^{-5}$
and $Latex$ 2 was diluted to $1.58\times 10^{-5}$. The solvent is the fresh
de-ionized water from a Milli-Q Plus water purification system (Millipore,
Bedford, with a 0.2 $\mu $m filter). Because the sizes of polystyrene latex
particles are small and\ the refractive index difference between the
polystyrene latex and the milli-Q water (refractive index 1.332) is large,
the m-1 is 0.19 and the ``phase shift'' $\frac{4\pi }{\lambda }R|m-1|$ $%
\left[ 4,10\right] $ is equal to 0.21 if the sample $Latex$ 2 is considered
and the refractive index 1.591 of polystyrene at 590 nm and $20^{o}C$ is
used, so the mono-disperse model $G\left( R_{s}\right) =\delta \left(
R_{s}-\left\langle R_{s}\right\rangle \right) $\ was used to obtain the
approximate values of the mean static radii $\left\langle R_{s}\right\rangle 
$ for the two commercial polystyrene latex samples respectively. The
commercial values of the mean radii and standard deviations of the two
samples shown in Table\ 1 were input Eq. \ref{main} to get $I_{s}/I_{inc}$
respectively. In order to compare with the experimental data, the calculated
and experimental values at $q=0.01887$ $nm^{-1}$\ were set equal. Figure 1.a
shows the results of $Latex$ 1. In order to compare the expected values $%
\left\langle R_{g}^{2}\right\rangle _{cal}^{1/2}$ with experimental values
of $\left\langle R_{g}^{2}\right\rangle _{Zimm}^{1/2}$, the commercial
values of the mean radii and standard deviations also\ were input Eq. \ref
{RG} respectively to obtain the expected root mean-square radii of gyration $%
\left\langle R_{g}^{2}\right\rangle _{cal}^{1/2}$. Meanwhile the root
mean-square radii of gyration were measured using the Zimm plot
respectively. Figure 1.b shows the fit results of $Latex$ 1, $%
Kc/R_{vv}=1.29\times 10^{-8}+3.11\times 10^{-6}q^{2}$. The DLS data of two
standard polystyrene latex samples were measured under the same conditions
as the SLS\ data respectively and the apparent hydrodynamic radii were
obtained using the cumulant method. In order to conveniently compare with
the values calculated using the size information obtained from the SLS data
and avoid the contradictions of $\mu _{2}$, the values of apparent
hydrodynamic radii were obtained using the first cumulant. For the two
standard polystyrene latex samples, the values of apparent hydrodynamic
radii at a scattering angle of 40$^{o}$ were chosen as the results obtained
using the DLS technique since the values of apparent hydrodynamic radius
almost do not depend on the scattering angle. All results are listed in
Table 1.

\begin{center}
$\overset{\text{Table 1The commerical size information, }\left\langle
R_{s}\right\rangle \text{, values of }\left\langle R_{g}^{2}\right\rangle
_{Zimm}^{1/2}\text{ , }\left\langle R_{g}^{2}\right\rangle _{cal}^{1/2}\text{
and }R_{h,app}\text{ at a scattering angle of 40}^{o}\text{.}}{
\begin{tabular}{|c|c|c|c|c|c|}
\hline
$\left\langle R\right\rangle \left( nm\right) \left( comm\right) $ & $\sigma
\left( nm\right) \left( comm\right) $ & $\left\langle R_{g}^{2}\right\rangle
^{1/2}\left( nm\right) $ & $\left\langle R_{g}^{2}\right\rangle
_{cal}^{1/2}\left( nm\right) $ & $\left\langle R_{s}\right\rangle \left(
nm\right) $ & $R_{h,app}\left( nm\right) $ \\ \hline
33.5$\left( Latex\text{ 1}\right) $ & 2.5 & 26.7 & 26.9 & 33.30$\pm $0.18 & 
37.27$\pm $0.09 \\ \hline
55$\left( Latex\text{ 2}\right) $ & 2.5 & 46.8 & 43.2 & 56.77$\pm $0.04 & 
64.48$\pm $0.56 \\ \hline
\end{tabular}
\ }$
\end{center}

For Poly($N$-isopropylacrylamide) $\left( PNIPAM\right) $ microgel samples,
the synthesis of gel particles was described elsewhere $\left[ 11,12\right] $%
. Equation \ref{main} was used to fit the data of the PNIPAM microgel sample
that the molar ratio of $N,N^{\prime }$-methylenebisacrylamide over\ $N$%
-isopropylacrylamide is $5\%$. The data were measured at a temperature of $%
29^{o}C$. The concentration is $8.38\times 10^{-6}$.\ The fit results at
different scattering vector ranges are shown in Table 2.

\begin{center}
$\overset{\text{Table 2 The fit results for PNIPAM microgel sample at
different scattering vector ranges and a temperature of 29}^{o}C\text{.}}{
\begin{tabular}{|c|c|c|c|}
\hline
$q\left( 10^{-3}nm^{-1}\right) $ & $\left\langle R_{s}\right\rangle (nm)$ & $%
\sigma (nm)$ & $\chi ^{2}$ \\ \hline
3.45 to 9.05 & 189.92$\pm $30.58 & 38.12$\pm $15.69 & 1.44 \\ \hline
3.45 to 11.2 & 199.50$\pm $10.45 & 32.84$\pm $6.49 & 1.17 \\ \hline
3.45 to 13.2 & 210.80$\pm $2.39 & 24.66$\pm $2.29 & 1.03 \\ \hline
3.45 to 14.2 & 215.47$\pm $1.47 & 19.91$\pm $1.87 & 1.07 \\ \hline
3.45 to 15.2 & 216.94$\pm $0.60 & 18.10$\pm $1.02 & 1.06 \\ \hline
3.45 to 16.1 & 216.69$\pm $0.40 & 18.46$\pm $0.71 & 1.07 \\ \hline
3.45 to 17.0 & 216.71$\pm $0.25 & 18.33$\pm $0.50 & 1.07 \\ \hline
3.45 to 17.9 & 217.55$\pm $0.14 & 16.73$\pm $0.36 & 1.23 \\ \hline
3.45 to 18.7 & 217.98$\pm $0.09 & 15.66$\pm $0.27 & 1.51 \\ \hline
3.45 to 19.5 & 218.14$\pm $0.06 & 14.57$\pm $0.17 & 2.49 \\ \hline
3.45 to 20.3 & 218.19$\pm $0.07 & 14.42$\pm $0.17 & 3.58 \\ \hline
\end{tabular}
}$
\end{center}

When Eq. \ref{main} was fit to the data, it was found that the results for
mean static radii $\left\langle R_{s}\right\rangle $ and standard deviation $%
\sigma $ depended on the scattering vector range being fit, as shown in
Table 2. If a small scattering vector range is chosen, the parameters are
not well-determined. As the scattering vector range is increased, $\chi ^{2}$
and the uncertainties in the parameters decrease and $\left\langle
R_{s}\right\rangle $ and $\sigma $ stabilize. If the fitting scattering
vector range continues to increase, the values of $\left\langle
R_{s}\right\rangle $ and $\sigma $ begin to change and $\chi ^{2}$ grows.
This is the results of the deviation between the experimental and
theoretical scattered light intensity in the vicinity of the scattered
intensity minimum. This minimum lies at about the scattering vector $0.0207$ 
$nm^{-1}$. In this range, most of the scattered light is cancelled due to
the light interference. So many other characteristics of particles can show
the effects on the scattered light intensity, for example: the particle
number distribution deviates from a Gaussian distribution, the particle
shape deviates from a perfect sphere and the density of particles deviates
from homogeneity, etc. In order to avoid the effects of light interference,
the stable fit results during the scattering vector range from $0.00345$ $%
nm^{-1}$ to $0.0170$ $nm^{-1}$ are chosen as the size information obtained
using the SLS technique. The size information is $\left\langle
R_{s}\right\rangle =216.7\pm 0.3$ $nm$, $\sigma =18.3\pm 0.5$ $nm$ and $\chi
^{2}=1.07$. The fit results in this range are shown in Fig. 2.

Using the standard method of cumulant, the apparent hydrodynamic radii can
be obtained at different scattering angles from $g^{\left( 2\right) }\left(
\tau \right) $ measured under the same conditions as the SLS data. The
values of apparent hydrodynamic radius $R_{h,app}$ are related to the
scattering angle. The value is about $280$ $nm$. Now using the light
scattering techniques, different particle sizes for a same particle system
can be obtained. This difference will not only influence the analysis of the
physical quantities that are related to the particle sizes, but also, more
importantly, it will bring one fundamental question: which size information
is the better approximation of the particle sizes? For the two commercial
polystyrene latex samples, the value obtained using the SLS technique is
consistent with that measured using TEM. Further the effects brought by this
difference to the standard method of cumulant also can be analyzed. From the
process that $g^{(2)}(\tau )$ is obtained, it is determined by both the
optical and hydrodynamic characteristics of particles. If the normalized
time auto-correlation function of the electric field of the scattered light $%
g^{(1)}(\tau )$ is written out in detail for homogeneous spherical particles
using the first cumulant, the following equation can be obtained

\begin{equation}
g^{\left( 1\right) }\left( \tau \right) =\exp \left( -q^{2}\left\langle
D\right\rangle \tau \right) =\frac{\int R_{s}^{6}G\left( R_{s}\right)
P\left( q,R_{s}\right) \exp \left( -q^{2}D\tau \right) dR_{s}}{\int
R_{s}^{6}P\left( q,R_{s}\right) G\left( R_{s}\right) dR_{s}},
\label{firstmom}
\end{equation}
where $D$ is the diffusion coefficient. From the Stokes-Einstein relation $D=%
\frac{k_{B}T}{6\pi \eta _{0}R_{h}}$ and $\left\langle D\right\rangle =\frac{%
k_{B}T}{6\pi \eta _{0}R_{h,app}},$ here $\eta _{0}$, $k_{B}$ and $T$ are the
viscosity of the solvent, Boltzmann's constant and absolute temperature,
respectively and $R_{h}$ is the hydrodynamic radius of a particle. $R_{h,app}
$ is the apparent hydrodynamic radius obtained using the cumulant method.

In Eq. \ref{firstmom}, the quantity $\exp \left( -q^{2}D\tau \right) $ is
related to the hydrodynamic characteristics while $R_{s}^{6}P\left(
q,R_{s}\right) $ is determined by the optical features of particles. As a
result, $g^{\left( 1\right) }\left( \tau \right) $ is determined by both the
optical and hydrodynamic features of particles but not purely determined by
one of them. For an approximate mono-disperse system, equation $g^{\left(
1\right) }\left( \tau \right) =\exp \left( -q^{2}D\tau \right) $ is a very
good method to obtain the hydrodynamic sizes of particles. When the
distribution of particles needs to be obtained, exactly the distribution
obtained using the cumulant method is a composite distribution $G(q,R_{h},f)$%
, here $f$ represents the relationship between the hydrodynamic radius $%
R_{h} $ and the static radius $R_{s}$. For narrow distributions, we ever
simply assumed that the relationship between the static and hydrodynamic
radii is $a=R_{h}/R_{s}$, here $a$ is a constant. With the size information
obtained using the SLS technique, the expected values of apparent
hydrodynamic radius can be obtained using Eq. \ref{firstmom}. The same
PNIPAM microgel sample was measured at a temperature of $40^{o}C$. The fit
results are $\left\langle R_{s}\right\rangle =139.3\pm 0.3$ $nm$, $\sigma
=12.4\pm 0.6$ $nm$ and $\chi ^{2}=5.50$. When the constant $a$ was chosen to
be 1.10, the five measured values obtained using the first cumulant and the
expected results are shown in Fig. 3. In order to avoid the consideration
for the large values of $\chi ^{2}$, all fit results are chosen under this
condition $\chi ^{2}\leq 2$.

The difference between the static and apparent hydrodynamic radii cannot be
totally explained by the effects of the different weights of particle sizes
and the scattering angle. For other samples, we have obtained the same
conclusion. The static radius and the apparent hydrodynamic radius are
different physical quantities.

Eq. \ref{main} provides a method to measure accurately the particle size
distribution and makes it possible to measure the average molar mass of
large particles. Comparing our method with the DLS technique, the simple
particle size $R_{s}$ and size distributions $G\left( R_{s}\right) $ of
dilute homogenous spherical particles can be directly obtained. The small
poly-dispersities that cannot be measured using the cumulant method have
been obtained. If the absolute magnitude of the scattered intensity and some
constants that are related to the instrument and samples are known, the
average molar mass for large particles can be measured. The apparent
hydrodynamic radius $R_{h,app}$ obtained using the cumulant method is a
composite size obtained from the average of $\exp \left( -q^{2}D\tau \right) 
$ in distribution $G\left( R_{s}\right) $ with the weight $R_{s}^{6}P\left(
q,R_{s}\right) $. The simple sizes and distributions obtained using the SLS
technique are the physical quantities that people really want to obtain from
the experimental data. They let us avoid the other parameters' effects when
the effects of particle sizes are analyzed.

Fig. 1 The experimental and expected values of $I_{s}/I_{inc}$ and the Zimm
plot analysis for $Latex$ 1. (a) The circles show the experimental data and
the line represents the expected values of $I_{s}/I_{inc}$. (b) The circles
show the experimental data and the line shows a linear fit to the plot of $%
Kc/R_{vv}$ as a function of $q^{2}$.

Fig. 2 The experimental and fit results for PNIPAM microgel sample at a
temperature of 29$^{o}C$. The circles show the experimental data, the line
shows the fit results and the diamonds show the residuals: $\left(
y_{i}-y_{fit}\right) /\sigma _{i}$.

Fig. 3 The experimental and expected values of apparent hydrodynamic radii
for the PNIPAM microgel sample at a temperature of $40^{o}C$. The circles
show the experimental data, the diamonds show the expected results
calculated using the size information obtained from SLS.

$\left[ 1\right] $ D. E. Koppel, J. Chem. Phys. 1972, 57, 4814

$\left[ 2\right] $ C. B. Bargeron, J. Chem. Phys. 1974, 61, 2134

$\left[ 3\right] $ J. C. Brown, P. N. Pusey and R. Dietz, J. Chem. Phys.
1975, 62, 1136

$\left[ 4\right] $ B. J. Berne and R. Pecora, Dynamic Light Scattering.
Robert E. Krieger Publishing Company, Malabar, Florida, 1990

$\left[ 5\right] $ B. J. Frisken, Appl. Opt. 2001, 40(24), 4087

$\left[ 6\right] $ B. H. Zimm, J. Chem. Phys. 1948, 16, 1099

$\left[ 7\right] $ P. N. Pusey and W. van Megen, J. Chem. Phys., 1984, 80,
3513

$\left[ 8\right] $ K. B. Strawbridge and F. R. Hallett, Macromolecules,
1994, 27, 2283

$\left[ 9\right] $ H. Schnablegger and O. Glatter, J. Colloid. Interface
Sci. 1993, 158, 228

$\left[ 10\right] $ H. C. van de Hulst, Light Scattering by Small Particles,
Dover Publications, Inc. New York, 1981

$\left[ 11\right] $ J. Gao and B. J. Frisken, Langmuir, 2003, 19, 5217

$\left[ 12\right] $ J. Gao and B. J. Frisken, Langmuir, 2003, 19, 5212

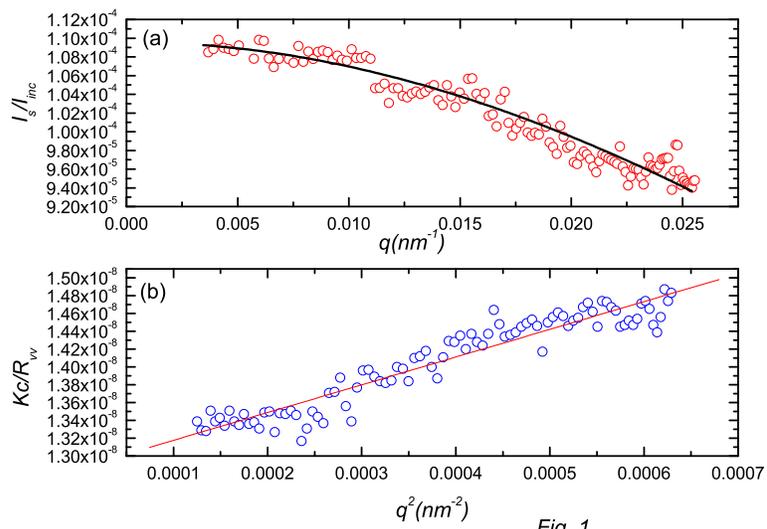
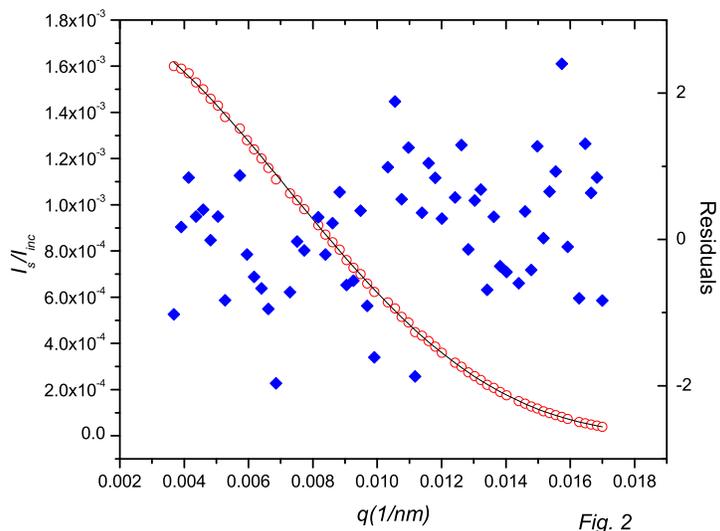
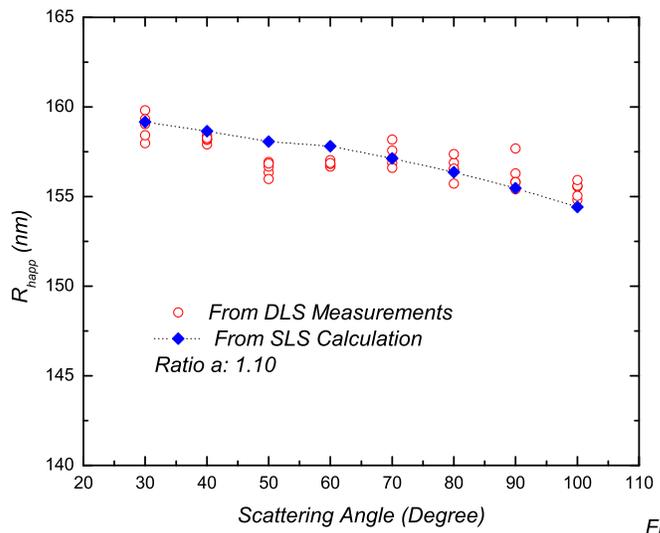
\end{document}